\documentstyle[prl,aps]{revtex}

\begin{document}
\author{Jian-Qi Shen \footnote{E-mail address: jqshen@coer.zju.edu.cn}}
\address{Zhejiang Institute of Modern Physics and Department of Physics,
\\Zhejiang University, Hangzhou 310027, P.R. China}
\date{\today }
\title{The self-induced charge current in the experiments detecting photon rest mass}
\maketitle

\begin{abstract}
This note considers two related topics regarding the {\it
self-induced charge current} in the experiments detecting photon
rest mass: (i) the effective rest mass, $(0.3\sim 1)\times
10^{-54}$ Kg, of photons arising from the ions in the secondary
cosmic ray is compared to Luo {\it et al.}'s new upper limit on
photon rest mass ($1.2\times 10^{-54}$ Kg); (ii) there are two
main classes of experiments to detect photon rest mass based
respectively on the Amp\`{e}re-Maxwell-Proca equation and the
Lagrangian (or Hamiltonian) density of electromagnetically
interacting system.
\\ \\

 PACS number(s): 03.50.De

\end{abstract}

During the last three decades, photon rest mass problem captured
special attention of many investigators who have reported several
experimental upper limits on the photon mass $m_{\gamma}$ by using
various
methods\cite{Feinberg,Williams,Goldhaber,Barrow,Davis,Gintsburg,Chernikov,Fischbach,Lakes}.
Since the radiation from pulsars and/or certain explosive
astrophysical events at high redshift can be used to place the
severe limits on both the fractional variation ($\triangle c/c$)
in the speed of light with frequency and the photon rest
mass\cite{Schaefer}, in 1969 Feinberg analyzed the data obtained
in the detection of the sharply defined radio and optical pulses
from pulsars and indicated that the rest mass of a real photon is
less than $10^{-47}$ Kg\cite{Feinberg}. Goldhaber {\it et al.}
discussed the method used to set a limit on the photon rest mass
({\it e.g.}, sensitive techniques which test Coulomb's law and its
analog in magnetostatics) and showed that the laboratory tests of
Coulomb's law\cite{Williams} gave an upper limit on photon rest
mass of $2\times 10^{-50}$ Kg and the constant ``external''
magnetic field experiment at the Earth's surface obtained a limit
$m_{\gamma}\leq 4\times 10^{-51}$ Kg\cite{Goldhaber}. The
experiments of Pioneer-10 measurement of Jovian magnetic
field\cite{Davis} and hydromagnetic wave in the Earth's
magnetosphere and solar wind\cite{Gintsburg} derived an upper
limit on $m_{\gamma}$ of about $10^{-52}$ Kg. In 1992 Chernokov
{\it et al.} suggested a new method to set an upper limit on the
photon mass at low temperature ($1.24$ K) based on a null test of
Amp\`{e}re's law and their experiment resulted in an upper limit
about $10^{-48}$ Kg\cite{Chernikov}. Fischbach {\it et al.}
investigated a new geometric limit on the photon rest mass which
was derived from an analysis of satellite measurements of the
Earth's magnetic field. The order of magnitude of the upper limit
on $m_{\gamma}$ obtained by Fischbach {\it et al.} was $10^{-51}$
Kg\cite{Fischbach}. In 1998, Lakes reported that their
experimental approach based on a toroid Cavendish balance which
was used to evaluate the product of photon rest mass squared and
the ambient cosmic magnetic vector potential gave an upper limit
on $m_{\gamma}$ of about $4\times 10^{-52}$ Kg\cite{Lakes}. More
recently, Luo {\it et al.} obtained the most new upper limit on
photon rest mass of $1.2\times 10^{-54}$ Kg by means of {\it
rotating torsion balance} experiment\cite{Luo}.

Although both Luo {\it et al.}'s experimental scheme and their
obtained results are excellent and impressive, here we have a
supplement to their results, since our evaluation shows that the
{\it photon effective rest mass} due to the {\it self-induced
charge currents}\cite{Ho,Scr} in the environmental dilute plasma
({\it e.g.}, the muon ($\mu$) component and alpha-particles in the
secondary cosmic ray flux) is just the {\it same} order of
magnitude\cite{Shen} as Luo's obtained upper limit on photon mass.

In addition, it is worthwhile to point out that there are two main
classes of experiments to detect photon rest mass based on the
Amp\`{e}re-Maxwell-Proca equation and the Lagrangian (or
Hamiltonian) density of electromagnetically interacting system,
respectively. The torsion balance experiments\cite{Lakes,Luo}
considered here fall into the latter class ({\it i.e.}, based on
the electromagnetic Lagrangian and Hamiltonian density).

In what follows we will discuss these two problems.

We consider the potential effects of self-induced charge currents
arising in the torsion balance experiments\cite{Lakes,Luo}. The
Lagrangian density of electrodynamics reads
\begin{equation}
{\mathcal
L}=-\frac{1}{4}F_{\mu\nu}F_{\mu\nu}-\frac{1}{2}\mu^{2}_{\gamma}A_{\mu}A_{\mu}+\mu_{0}J_{\mu}A_{\mu},
\label{eqn1}
\end{equation}
where
$\mu_{\gamma}^{2}=\left({\frac{m_{\gamma}c}{\hbar}}\right)^{2}$
with $m_{\gamma}$, $\hbar$ and $c$ being the photon rest mass,
Planck's constant and speed of light in a free vacuum,
respectively, and $\mu_{0}$ denotes the magnetic permeability in a
vacuum. Here the summation over the repeated indices is implied.
It follows from (\ref{eqn1}) that the canonical momentum density
of electromagnetic field reads $\pi_{\mu}=\frac{\partial{\mathcal
L}}{\partial\dot{A}_{\mu}}$ or $\vec{\pi}(x)=-{\bf E}(x)$. In the
ions plasma, due to the conservation law of canonical momentum
density of interacting electromagnetic system, {\it i.e.},
$\frac{{\rm d}}{{\rm d}t}(m{\bf v}+e{\bf A})=0$, we have ${\bf
v}=-\frac{e}{m}{\bf A}+{\bf C}$ with ${\bf C}$ being a constant
velocity. By using the formula ${\mathcal H}=-{\bf
E}\cdot{\dot{\bf A}}-{\mathcal L}$ for the system Hamiltonian and
the electric current density ${\bf J}=Ne{\bf
v}=-\frac{N{e}^{2}}{m}{\bf A}+Ne{\bf C}$ with $N$ standing for the
volume density of ions , one can arrive at (in SI)
\begin{equation}
{\mathcal H}=\frac{1}{2\mu_{0}}\left[{\frac{{\bf
E}^{2}}{c^{2}}+{\bf
B}^{2}+\left(\mu_{\gamma}^{2}+2\frac{\mu_{0}N{e}^{2}}{m}\right){\bf
A}^{2}+\frac{1}{\mu_{\gamma}^{2}c^{2}}\left(\nabla\cdot{\bf
E}-\frac{{\bf J}_{0}}{\epsilon_{0}}\right)^{2}}\right]+{\bf
J}_{0}{\bf A}_{0}-Ne{\bf C}\cdot{\bf A}, \label{eqn3}
\end{equation}
where ${\bf J}_{0}=\rho$ and ${\bf A}_{0}=\phi$ are respectively
the electric charge density and electric scalar potential, and
$\epsilon_{0}$ represents the electric permittivity in a vacuum
and $c=\frac{1}{\sqrt{\epsilon_{0}\mu_{0}}}$. Note that here
$\frac{1}{\mu_{\gamma}^{2}c^{2}}\left(\nabla\cdot{\bf
E}-\frac{{\bf J}_{0}}{\epsilon_{0}}\right)^{2}$ results from both
the mass term ${\frac{\mu_{\gamma}^{2}}{c^{2}}}{\bf A}_{0}^{2}$
and the Gauss's law $\nabla\cdot{\bf E}=-\mu_{\gamma}^{2}{\bf
A}_{0}+\frac{{\bf J}_{0}}{\epsilon_{0}}$.

It follows from (\ref{eqn3}) that the {\it total effective rest
mass squared} of electromagnetic fields in ions plasma is given as
follows
\begin{equation}
\mu_{\rm tot}^{2}=\mu_{\gamma}^{2}+2\frac{\mu_{0}N{e}^{2}}{m}.
\label{eqn4}
\end{equation}

It is known that at the sea level, the current density of muon
component in secondary cosmic rays is about $1\times 10^{-2}{\rm
cm}^{-2}\cdot{\rm s}^{-1}$\cite{Mei}. Assuming that the muon
velocity approaches the speed of light, the volume density of muon
can be derived and the result is $N_{\mu}=0.3\times 10^{-6}$ ${\rm
m}^{-3}$. So, according to the formula (\ref{eqn4}), the photon
effective rest mass due to the {\it self-induced muon charge
current} takes the form

\begin{equation}
m_{\rm
eff}=\frac{\hbar}{c^{2}}\sqrt{\frac{2N_{\mu}e^{2}}{\epsilon_{0}m_{\mu}}}
\label{eq m1}
\end{equation}
with $m_{\mu}$ denoting the muon mass, electromagnetic wave with
wavelength $\lambda \gg 100$ m\cite{Shen} at the sea level
acquires an effective rest mass about $0.3\times 10^{-53}$ Kg.
Hence, likewise, in the {\it rotating torsion balance} experiment
the ambient cosmic magnetic vector potentials (interstellar
magnetic fields) with low or zero frequencies may also acquire
such effective rest mass.

Note that the above-mentioned muon volume density ($0.3\times
10^{-6}$ ${\rm m}^{-3}$) is the datum obtained only at the sea
level. In order to evaluate the muon volume density in the
environment where Luo's experiment was performed, readers may be
referred to the following handbook data of muon current density in
underground cosmic rays: the muon current densities are
respectively $10^{-4}$ ${\rm cm}^{-2}\cdot{\rm s}^{-1}$ and
$10^{-6}$ ${\rm cm}^{-2}\cdot{\rm s}^{-1}$ at the equivalent water
depths of $100$ m and $1000$ m under the ground\cite{Mei}. Since
in Luo's experiment the total apparatus is located in a cave
laboratory, on which the least thickness of the cover is more than
$40$ m\cite{Luo}, it is reasonably believed that the muon density
of secondary cosmic rays in the vacuum chamber of Luo's experiment
may be one or two orders of magnitude less than that at the sea
level. This, therefore, means that by using the formula (\ref{eq
m1}) the effective rest mass acquired by photons in Luo's vacuum
chamber may be about $(0.3\sim 1)\times 10^{-54}$ Kg which can be
compared to Luo's result ($1.2\times 10^{-54}$ Kg).

It is believed that the air molecules in the low-pressure vacuum
chamber of torsion balance experiment cannot be easily ionized by
the alpha-particles, electrons and protons of cosmic rays,
because, for example, it is readily verified that the mean free
path of alpha-particle moving at about $10^{7}{\rm m}\cdot{\rm
s}^{-1}$ in the dilute air with the pressure being only $10^{-2}$
Pa\cite{Luo} is too large ({\it i.e.}, more than $10^{6}$
m)\cite{Shen} for the dilute air molecules to be ionized. So, the
air medium in the low-pressure vacuum chamber has nothing to
contribute to the effective rest mass of photons. However, as far
as the alpha-particle, $\pi$ mesons, electrons {\it etc.} in
secondary cosmic rays are concerned, these ions will also
contribute about $10^{-54}$ Kg to photon rest mass, since, for
instance, the volume density of alpha-particles and other ions at
the sea level in secondary cosmic rays is about one order of
magnitude less than that of muons\cite{Mei}.

Thus the photon effective rest mass resulting from both muons and
other ions in secondary cosmic rays can be compared to the newly
obtained upper limit ($1.2\times 10^{-54}$ Kg) by Luo {\it et al.}
in their recent rotating torsion balance experiment\cite{Luo}. For
this reason, Luo's experimental upper limit on photon mass is said
to be just a critical value. It is reasonably believed that Luo's
recent experimental result is of physical interest. However, the
photon mass problem still deserves further experimental
investigation so as to improve the present upper limit on photon
mass. It is claimed that, in the near future, only the
contribution of ions in environmental dilute plasma, {\it i.e.},
the secondary cosmic rays is ruled out, can we deal better with
the experimental schemes, which can place severe limit on the
photon rest (intrinsic) mass and other related experimental
results. In addition, it is apparently seen that not only the
photon intrinsic rest mass but also the effective mass due to
media dispersion will play important roles in both fundamental and
applied physics. For example, taking into account the influence of
the media dispersion and the consequent photon effective mass on
the geometric phases of photons propagating inside a noncoplanarly
curved fiber\cite{Ma} is physically interesting, which is under
consideration and would be published elsewhere.
\\ \\

Additionally, the difference between the two categories of
$\mu_{\rm eff}^{2}$ based on different physical mechanisms should
be pointed out: the theoretical mechanism of torsion balance
experiment lies in the Lagrangian or Hamiltonian (\ref{eqn3}) of
interacting electromagnetic system, so, the effective rest mass
squared of electromagnetic fields is $\mu_{\rm
eff}^{2}=2\frac{\mu_{0}N{e}^{2}}{m}$. However, for those
experimental schemes\cite{Feinberg,Chernikov,Fischbach} of photon
rest mass stemming from the Amp\`{e}re-Maxwell-Proca equation
$\nabla\times{\bf B}=\mu_{0}\epsilon_{0}\frac{\partial}{\partial t
}{\bf E}+\mu_{0}{\bf J}-\mu^{2}_{\gamma}{\bf A}$ and the
consequent
\begin{equation}
\nabla\times{\bf B}=\mu_{0}\epsilon_{0}\frac{\partial}{\partial t
}{\bf E}-[\mu^{2}_{\gamma}+\frac{\mu_{0}N{e}^{2}}{m}]{\bf
A}+\mu_{0}Ne{\bf C},                     \label{eqmaxwell}
\end{equation}
the total mass squared $\mu_{\rm tot}^{2}$ should be $\mu_{\rm
tot}^{2}=\mu_{\gamma}^{2}+\frac{\mu_{0}N{e}^{2}}{m}$ rather than
that in (\ref{eqn4}). This minor difference between these two
$\mu_{\rm tot}^{2}$ results from the derivative procedure applied
to the Lagrangian density by using the Euler-Lagrange equation.
Thus the effective rest mass squared of electromagnetic fields in
the experiments based on the Amp\`{e}re-Maxwell-Proca equation
(\ref{eqmaxwell}) is $\mu_{\rm eff}^{2}=\frac{\mu_{0}N{e}^{2}}{m}$
rather than $\mu_{\rm eff}^{2}=2\frac{\mu_{0}N{e}^{2}}{m}$. The
so-called experimental realizations stemming from the
Amp\`{e}re-Maxwell-Proca equation are as follows: pulsar test of a
variation of the speed of light with frequency\cite{Feinberg};
geomagnetic limit on photon mass based on the analysis of
satellite measurements of the Earth's field\cite{Fischbach};
experimental test of Amp\`{e}re's law at low
temperature\cite{Chernikov}, {\it etc.}. In all these experimental
schemes, the effective mass squared is $\mu_{\rm
eff}^{2}=\frac{\mu_{0}N{e}^{2}}{m}$.

It is shown that in the rotating torsion balance experiments the
extra torque $2\frac{\mu_{0}Ne^{2}}{m}{\bf a}_{d}\times {\bf A}$
produced by the interaction of magnetic dipole vector potential
moment ${\bf a}_{d}$ with the ambient cosmic magnetic vector
potential will also acts upon the toroid in the torsion balance
experiments\cite{Lakes,Luo} whose apparatus is immersed in the
dilute plasma such as the cosmic rays and radon gas. Here the
photon effective rest mass squared is $\mu_{\rm
eff}^{2}=2\frac{\mu_{0}N{e}^{2}}{m}$ rather than $\mu_{\rm
eff}^{2}=\frac{\mu_{0}N{e}^{2}}{m}$, as stated above.
\\ \\

\textbf{Acknowledgements}  This project was supported partially by
the National Natural Science Foundation of China under the project
No. $90101024$.

\end{document}